\documentclass[aps,pra,twocolumn,superscriptaddress,showpacs,longbibliography]{revtex4-2}
\usepackage{amsmath,amssymb,amsfonts,bbm,float,graphics,epsfig,epstopdf,color,verbatim,tabularx,bm,multirow,hyperref,ulem,times,subfigure}
\hypersetup{
  colorlinks, linkcolor={blue},
  citecolor={blue}, urlcolor={blue}
}

\begin{document}
\title{Emergent synchronization mode in coupled Rydberg atomic chains}

\author{Weilun Jiang}
\email{wljiang@sxu.edu.cn}
\affiliation{State Key Laboratory of Quantum Optics Technologies and Devices, Institute of Opto-Electronics, Collaborative Innovation Center of Extreme Optics, Shanxi University, Taiyuan, 030006, China }

\begin{abstract}
We report a new oscillatory form in the two coupled dissipative Rydberg atomic chains by modulating its spacing. Such oscillation has $\pi$-phase difference between two neighboring sites, which distinguishes itself from antiferromagnetic-type synchronization in the previous studies. Theoretically, we find a phase with coexisting two types of continuous time crystals, and recognize that the transition belongs to Hopf and pitchfork bifurcation. Furthermore, we generalize the conclusion to multiple chains and verify the uniqueness of the new synchronization mode. We also discuss its experimental feasibility.
\end{abstract}

\date{\today}
\maketitle

\section{Introduction}

Rydberg atoms are renowned for their long-range nonlinear interactions, making them an excellent experimental platform for studying many-body dynamics and performing quantum simulations \cite{Rydberg2010Weimer}. Theoretically, the inclusion of nonlinear interactions between highly excited atomic states renders the system’s equation of motion (EOM) analytically intractable and leads to rich and diverse dynamical behavior, and often giving rise to emergent phases of matter \cite{Tunable2016Labuhn,Observation2019Sylvain,Quantum2019Keesling,Quantum2021Samajdar,
bernienProbingManybodyDynamics2017,bluvsteinControllingQuantumManybody2021,bluvsteinQuantumProcessorBased2022,bornetScalableSpinSqueezing2023,gonzalez2025observation,chenContinuousSymmetryBreaking2023,chenSpectroscopyElementaryExcitations2025,deleseleucObservationSymmetryprotectedTopological2019,labuhnTunableTwodimensionalArrays2016,ebadiQuantumOptimizationMaximuma,ebadiQuantumPhasesMatter2021,ferioliNonequilibriumSuperradiantPhase2023,liangObservationThreephotonBound2018,manovitzQuantumCoarseningCollective2025,maskaraProgrammableSimulationsMolecules2025,omranGenerationManipulationSchrodinger2019,qiaoRealizationDopedQuantum2025a,schollQuantumSimulation2D2021,semeghini2021probing,Many2020Browaeys,Towards2018Nguyen}. Among them is an important class of non-equilibrium phases — continuous time crystals \cite{Quantum2012Wilczek,Zaletel2023Quantum,Kyprianidis2021Observation,Kessler2021Observation}, characterized by the spontaneous breaking of time-translation symmetry in the system's evolution.

Early theoretical proposals demonstrated that \cite{Iemini2018Boundary,Classical2020Yao}, in the presence of spontaneous emission, the interplay between driving and dissipation could induce persistent temporal oscillations, which were interpreted as signatures of time-crystalline behavior. Subsequently, these findings were verified both in various experimental system and theoretical model \cite{chitra2015dynamical,Gong2018Discrete,Buca2019Non-Stationary,zhu2019dickea,gambetta2019DiscreteTimeCrystals,Booker2020, alaeian2021limit}, termed as the dissipative time crystal. Among them, in hot atomic vapor experiments, luxuriant oscillatory behaviors have been observed through electromagnetically induced transparency (EIT) \cite{Emergence2023Wadenpfuhl,Dissipative2024Wu,Higher2024Liu,Ergodicity2024Ding,Bifurcation2025Liu,Jiao2025Observation}. Inspired by these developments, our work aims to explore novel oscillatory phases of matter by means of a simpler and more experimentally accessible design. 

In this work, we propose a system with two coupled and identical Rydberg atomic chains with adjustable spacing between them. Under appropriate parameters, we reproduce oscillatory modes with the same Rydberg population among alternate lattice sites, similar to those previously reported \cite{Antiferromagnetic2011Lee}. Remarkably, by tuning the inter-chain spacing, we also uncover a new class of stable oscillatory patterns characterized by on-site locking between the two chains, indicating a novel form of synchronization dynamics arising from weak inter-chain coupling. Our theoretical simulations reveal the coexistence of two types of continuous time crystals, each corresponding to a stable oscillatory solution. We further investigate its associated phase transitions and the feature of these oscillatory modes, providing a comprehensive understanding of the underlying dynamical behavior.

\section{Model and methods}

\begin{figure}[t]
  \centering
  \includegraphics[width=\columnwidth]{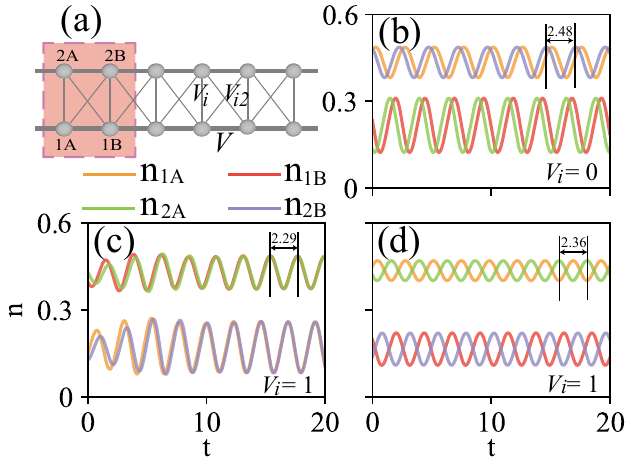}
  \caption{ (a) Sketch map of two coupled Rydberg chains. The region colored by red is the specified unit cell in the MF calculation, and the lattice sites are denoted as $1A, 1B, 2A, 2B$. The interaction considered in Eq. \eqref{eq4} are divided into three class $V$, $V_i$ and $V_{i2}$, which are represented by bonds with different thicknesses. (b) Time-evolution of Rydberg populations at decoupled case $V_i=0$. Two oscillations of both chains have same period but an arbitrary phase difference, determined by its initial conditions. (c,d) Time-evolution at $V_i = 1$ for different initial conditions. (c) shows the traditional AF-type oscillation pattern, and (d) shows the emergent $\text{AF}_2$-type oscillation. }
  \label{fig1}
\end{figure}

We consider two coupled Rydberg atomic chains. By design, the two chains have the same length and are aligned [Fig. \ref{fig1}(a)]. Here, two distances of the whole system are experimentally adjustable, the intra-chain and inter-chain length between two adjacent atoms. Therefore, we denote the two tunable parameters, the intra-chain interaction as $V$ and inter-chain interaction as $V_i$ to describe the nonlinear interaction between Rydberg excitations, and have $V \geq V_i$. The system Hamiltonian writes,
\begin{equation}
    \begin{aligned}
        H &= \frac{\Omega}{2} \sum_i ( \sigma_{1i}^x + \sigma_{2i}^x ) - \Delta \sum_i ( n_{1i} + n_{2i} ) \\
        &+ V \sum_{\langle i, j\rangle} ( n_{1i} n_{1j} + n_{2i} n_{2j})  + V_{i} \sum_i n_{1i} n_{2i} \\ 
         & + V_{i2} \sum_{\langle i, j\rangle}( n_{1i} n_{2j} + n_{2i} n_{1j}) + \cdots,
    \end{aligned}
\end{equation}
where the lattice coordinates of two chains are $1i$ and $2i$, respectively. $\langle \cdots \rangle$ denotes the nearest-neighbor sites. The other long-range interaction terms are absorbed in the ellipsis. 

We also introduce the spontaneous emission from the Rydberg level, denoted as $\gamma$. In the presence of weak Markovian dissipation, the dynamic of the system is characterized by the Master equation. The Lindblad operator is now $L = \sigma^-$, and the density matrix $\rho$ satisfies 
\begin{equation}
    \begin{aligned}
        \dot \rho &= -i \left[ H, \rho\right] + \mathcal{L}(\rho) \\
        \mathcal{L}(\rho) &= \gamma \sum_i \left( \sigma^-_i \rho \sigma^+_i -\frac{1}{2} \{ n_i, \rho\} \right).
    \end{aligned}
\end{equation}

We first employ the mean-field (MF) approach to deal with the problem. Considering the uniform solution, we denote $n = \langle n_i \rangle$ and $\sigma = \langle \sigma^-_i \rangle$. The EOM for the observable of each site on the single chain reads (See Appendix for the details),
\begin{equation}
    \begin{aligned}
        \partial_t n &= \Omega \text{Im} \sigma - \gamma n \\
        \partial_t \sigma &= -\frac{\gamma}{2} \sigma - i(\Delta - 2Vn) \sigma - i\frac{\Omega}{2} (2n-1).
    \end{aligned}
    \label{eq3}
\end{equation}
Here we only consider the nearest interaction, which contributes $2V$ to the Eq. \eqref{eq3} (See Appendix for the discussion on the next-nearest neighbor interaction). Ref. \cite{Antiferromagnetic2011Lee} has reported the fruitful analysis on the solution by the MF method, that the system shows an antiferromagnetic (AF) phase and periodic oscillation of the Rydberg population, i.e. $n_1 = n_3 = \cdots \neq n_2 = n_4 \cdots$. This results from the statement that the AF-type perturbation serves the lowest-order modification of the solution, and a fair approach is to introduce the sublattice $A$ and $B$ in the MF equation. Generalized to our lattice system, when taking $V_i = V$, the situation returns to that in Ref. \cite{Antiferromagnetic2011Lee}, which is based on the square lattice with the uniform Rydberg interaction. However, in consideration of the arbitrary value of $V_i$, we should include simultaneously the observables with respect to both two chains. That gives rise to the coupling equation of 4-sites unit cell in Eq. \eqref{eq4}. 

Since the Rydberg interaction strength decays as $\sim \frac{1}{r^6}$, where $r$ is the distance between two atoms, we only consider the several nearest-neighbor interaction $V, V_i, V_{i2}$ for simplication [Fig. \ref{fig1}(a)]. We also notice that $V_{i2}$ is fully determined by $V$ and  $V_{i}$. Under such parameter choice, the EOM of Eq. \eqref{eq4} is numerically solved using the Runge-Kutta method. For the parameter choice, to form the stable oscillatory solution, we choose $\Omega = 2.2, \Delta = 2.5, V = 5$ and all the parameters is in the unit of $\gamma $ \cite{Antiferromagnetic2011Lee}. In particular, for $^{87}\text{Rb}$, $\gamma = 6 \times 2\pi \text{MHz}$ \cite{Emergence2023Wadenpfuhl}, and for Cesium atoms, $\gamma = 5.2 \times 2\pi \text{MHz}$ \cite{Jiao2025Observation}. In the following content, we display the time-dependent solutions of Eq. \eqref{eq4}, accompanied with the stabilization analysis. 

\begin{widetext}
\begin{equation}
    \begin{aligned}
        \partial_t n_{1A} &= \Omega \text{Im} \sigma_{1A} - \gamma n_{1A}, \quad \partial_t \sigma_{1A} =  -\frac{\gamma}{2}  \sigma_{1A} - i(\Delta - 2Vn_{1B} - V_i n_{2A} -  V_{i2} n_{2B}) \sigma_{1A} - i\frac{\Omega}{2} (2n_{1A}-1) \\
        \partial_t n_{1B} &= \Omega \text{Im} \sigma_{1B} - \gamma n_{1B}, \quad \partial_t \sigma_{1B} = -\frac{\gamma}{2} \sigma_{1B} - i(\Delta - 2Vn_{1A} - V_i n_{2B} -  V_{i2} n_{2A} ) \sigma_{1B} - i\frac{\Omega}{2} (2n_{1B}-1) \\
        \partial_t n_{2A} &= \Omega \text{Im} \sigma_{2A} - \gamma n_{2A}, \quad \partial_t \sigma_{2A} = -\frac{\gamma}{2} \sigma_{2A} - i(\Delta - 2Vn_{2B} - V_i n_{1A} -  V_{i2} n_{1B}) \sigma_{2A} - i\frac{\Omega}{2} (2n_{2A}-1) \\
        \partial_t n_{2B} &= \Omega \text{Im} \sigma_{2B} - \gamma n_{2B}, \quad \partial_t \sigma_{2B} = -\frac{\gamma}{2} \sigma_{2B} - i(\Delta - 2Vn_{2A} - V_i n_{1B} -  V_{i2} n_{1A} ) \sigma_{2B} - i\frac{\Omega}{2} (2n_{2B}-1). \\                               
    \end{aligned}
    \label{eq4}
\end{equation}
\end{widetext}

\section{Main results}
In this section, we present our main results. The time-dependent evolution of Rydberg populations of 4-sites are shown in Fig. \ref{fig1}. Without $V_i$, the two chains have their own oscillations with the same frequency, while the phase is independent of another [Fig. \ref{fig1}(b)]. When we take $V_i = V$, there is only one oscillatory solution with the same population of $n_{1A}$ and $n_{2B}$ ($n_{2A}$ and $n_{1B}$). The solution is the so-called AF-type, and could be expressed as the phase coincidence of the two chains [Fig. \ref{fig1}(c)]. At intermediate $V_i$, we observe an emergent oscillatory mode with the same oscillatory amplitude of $n_{1A}$ and $n_{1B}$ [Fig. \ref{fig1}(d)]. Meanwhile, the oscillatory phase of two chains differs $\pi$, for which reason we denote it as the $\text{AF}_2$-type oscillation. This new emergent phase roots in the weaker coupling between two chains, which distinguishes its oscillatory pattern from the original behavior. In the following, we give a detailed theoretical analysis on such the phenomenon.   

\begin{figure*}[t]
  \centering
  \includegraphics[width=\textwidth]{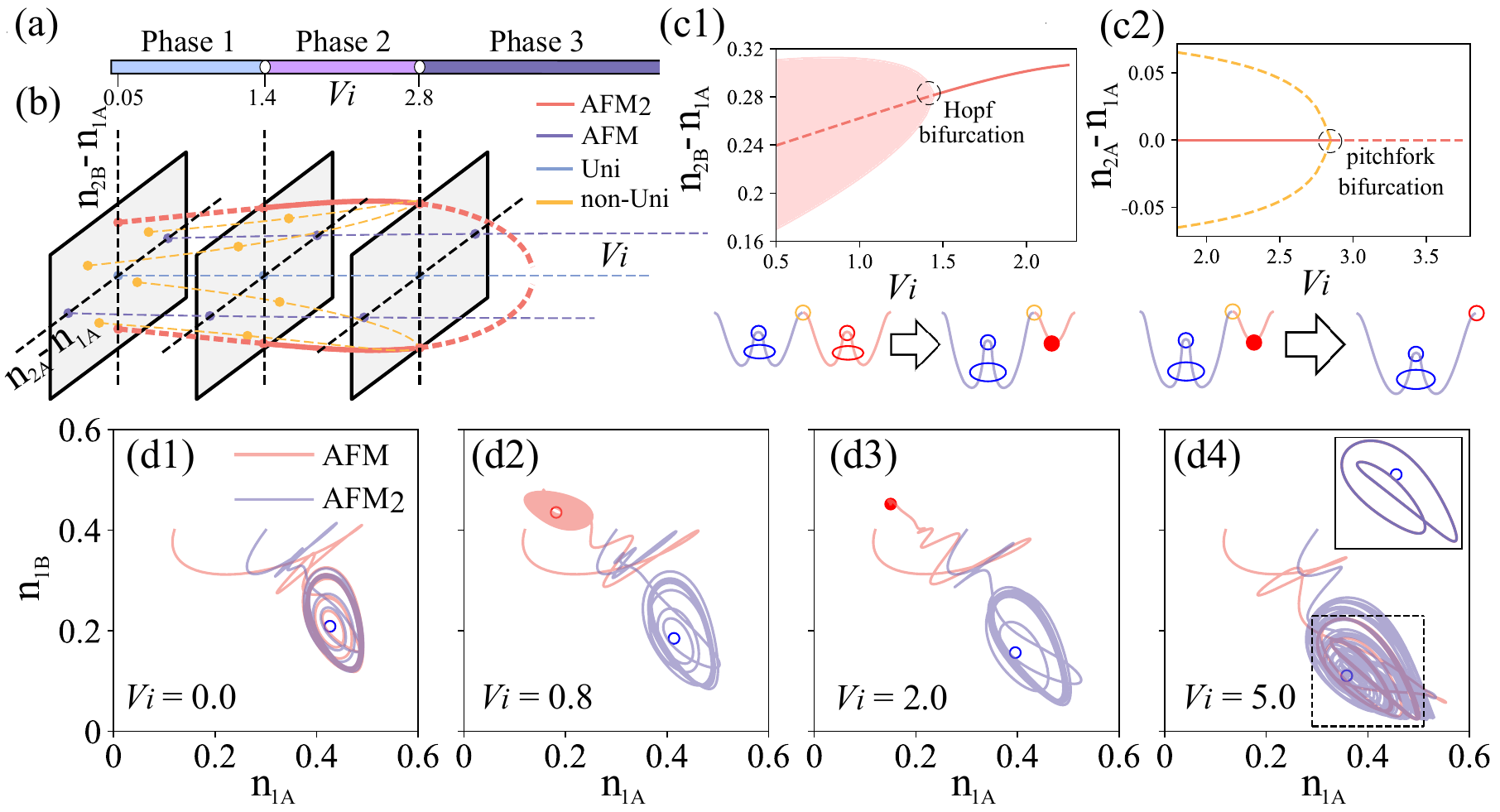}
  \caption{ (a) The phase diagram of two coupled Rydberg chains with respect to the inter-chain interaction $V_i$. The blue, pink and purple regions denote phase 1, 2, 3, respectively. The transition point are labeled by white circle. (b) Time-evolution of Rydberg population for different phases over a sufficiently long time. The red and blue lines represents two different initial conditions, corresponding two coexisting and distinguishable trajectories, which flows to AF-type and $\text{AF}_2$-type synchronization modes, respectively. The solid (hollow) circle is the associated stable (unstable) solution. The blue limit cycle (AF-type) is always existing, and encircles its unstable solution. The red limit cycle ($\text{AF}_2$-type) only exists in (b2). Note that (b1) shows the decoupled case, where only AF-type limit cycle survives. The sub-figure in (b4) is a clearer presentation of the limit cycle. (c) Stabilization analysis by varying $V_i$ in three-dimensional diagram. We begin at small $V_i = 0.05$ and end at $V_i = V = 5$. We select $V_i$ values at transition points, and the solutions are represented by the dots on the cross section. (d) Stabilization analysis showing the behavior near transition points in two-dimensional diagram and its sketch map. In the sketch map, the circles are the solutions from stabilization analysis, and the ovals indicate the oscillatory solutions. The wells denote the BOAs with associated colors. Note in (d1), the region covered by red is the oscillatory range for different $V_i$.}
  \label{fig2}
\end{figure*}

\section{Theoretical analysis}
We numerically solve the solution of Eq. \eqref{eq4} with all equations equal to zero, and perform the stabilization analysis for each solution. In practice, we set $V_{i2} = 0$ to reduce computational complexity. The results are displayed in the three-dimensional coordinate system. We classify all solutions to be four types, and conclude that in the Tab. \ref{tab1}. 

\begin{table}[h]
\centering
\caption{\bf Four different types of the steady solutions}
\begin{tabular}{cccc}
\hline
Type & Condition & Num. & Stability \\
\hline
uniform & $n_{1A} = n_{1B} = n_{2A} = n_{2B}$ & 1 & unstable \\
AF & $n_{1A} = n_{2B} \neq n_{1B} = n_{2A}$ & 2 & unstable \\
$\text{AF}_2$ & $n_{1A} = n_{2A} \neq n_{1B} = n_{2B}$ & 0/2 & unstable/stable \\
non-uniform & $n_{1A} \neq n_{2A} \neq n_{1B} \neq n_{2B}$ & 0/4 & unstable \\
\hline
\end{tabular}
\label{tab1}
\end{table}

Based on the number and stability of the $\text{AF}_2$ solutions, we could identify three different regions in the phase diagram with respect to $V_i$ [Fig. \ref{fig2}(a, b)]. In particular, we focus the solutions for the stability analysis near the phase transition point [Fig. \ref{fig2}(c)] for further analysis. The associated time evolution behavior with different initial states and display its trajectory in the parameter space are also displayed [Fig. \ref{fig2}(d)]. Now, we are ready to reveal the features of each phase and the type of transition points.

\textbf{Phase 0:}  Firstly, without the interaction between two chains, the oscillatory periods and amplitudes of each chain are the same [Fig. \ref{fig1}(b)]. This periodic structure of the solution in the phase space can also be interpreted as the limit cycle [Fig. \ref{fig1}(d1)] \cite{Colin2006Limit}, and the adjacent trajectories will finally flow to the cycle. It should be noted that although the trajectories of two limit cycles are degenerate, their associated phases are irrelevant and only determined by the initial condition. 

\textbf{Phase 1: }  Once we turn on $V_i$ for $V_i \lesssim 1.4$, the two chains are coupled with each other, and now a new synchronization mode occurs. We observe two coexisting stable limit cycles, that are similar to the bistable structure. Different initial conditions could lead to incompatible trajectories [Fig. \ref{fig2}(d2)]. Compared with the AF-type oscillatory pattern, the $\text{AF}_2$-type oscillation could also be categorized as a special type of AF-type oscillation. For the former, the AF correlations are along both inter-chain and intra-chain, thus constitute a checkboard pattern for instantaneous observation. While for the latter, the inter-chain AF correlation is now embodied in the phase correlation for the oscillation. 

\textbf{Phase 2: }  At large $V_i$ for $ 1.4 \lesssim V_i \lesssim 2.8$, the $\text{AF}_2$-type mode is now a stable solution, corresponding to a single attraction point in the parameter space [Fig. \ref{fig2}(d3)]. At this time, the stable solution owns only AF correlation along the intra-chain direction, and becomes uniform along the inter-chain direction.

\textbf{Phase 3: }  For $ 2.8 \lesssim V_i \leq V$, only stable AF-type oscillatory solution survives, which is consistent with the conclusion in Ref. \cite{Antiferromagnetic2011Lee} [Fig. \ref{fig2}(d4)]. 

From the perspective of the stabilization analysis [Fig. \ref{fig2}(b)], we find that the uniform solution with all equivalent Rydberg populations is always unstable. The reason is that such a solution becomes unstable by adding even tiny AF-type perturbation, and transfers to the stable AF-type oscillation. Here, the stable oscillation is bound to an unstable AF solution (colored blue), and the parameter points with respect to the unstable solutions are always enveloped with the limit cycle in the parameter space. Along this line, we identify that a new type of unstable solution exists in our model in phase 2 (red lines), which is the signal for the new oscillatory pattern. 

We proceed to analyze the transition between different phases. There are three transition points, where the first one is special since it represents the transition from infinitely many solutions to finitely many solutions. In this work, we focus on the other two as follows. 

\textbf{Transition between phase 1 and 2: }  We show the results of the stabilization analysis with respect to the $\text{AF}_2$-type axis, and the variation of the oscillation amplitude versus $V_i$ [Fig. \ref{fig2}(c1)]. Despite the presence of the AF-type limit cycle, the transition links two phases with a limit cycle and a stable solution. Mathematically, it is interpreted as the \textit{Hopf bifurcation} \cite{Hansjörg2012Bifurcation}, with the formation of the oscillatory solution. This can also be demonstrated in the sketch map of the parameter space, corresponding to the process that the maximum value in the $\text{AF}_2$ potential well decreases to be non-existence.   

\textbf{Transition between phase 2 and 3: } From the stabilization analysis results [Fig. \ref{fig2}(c2)], we numerically find the stable $\text{AF}_2$-type solution becomes unstable and the non-uniform solutions (colored by yellow) disappear at this transition point. We clearly show that the non-uniformity gradually decreases to zero. Mathematically, it corresponds to a  \textit{subcritical pitchfork bifurcation} \cite{Hansjörg2012Bifurcation}, where two unstable and one stable solutions collapse to one unstable solution. In the parameter space, the coexistence of two limit cycles (or stable solution) leads to two basis of attractions (BOAs) depicted by wells, labeled AF and $\text{AF}_2$. Both the uniform and non-uniform unstable solutions are located on the boundary of two BOAs. As $V_i$ increases, the area of $\text{AF}_2$-type BOA shrinks and finally becomes a single maximum point in AF-type BOA.  

Note at $V_i \sim 3.9$, the discrepancy for two unstable $\text{AF}_2$-type solutions vanishes, and finally they converge to the uniform unstable solution.  

\section{Extended models}

\begin{figure}[h]
  \centering
  \includegraphics[width=\columnwidth]{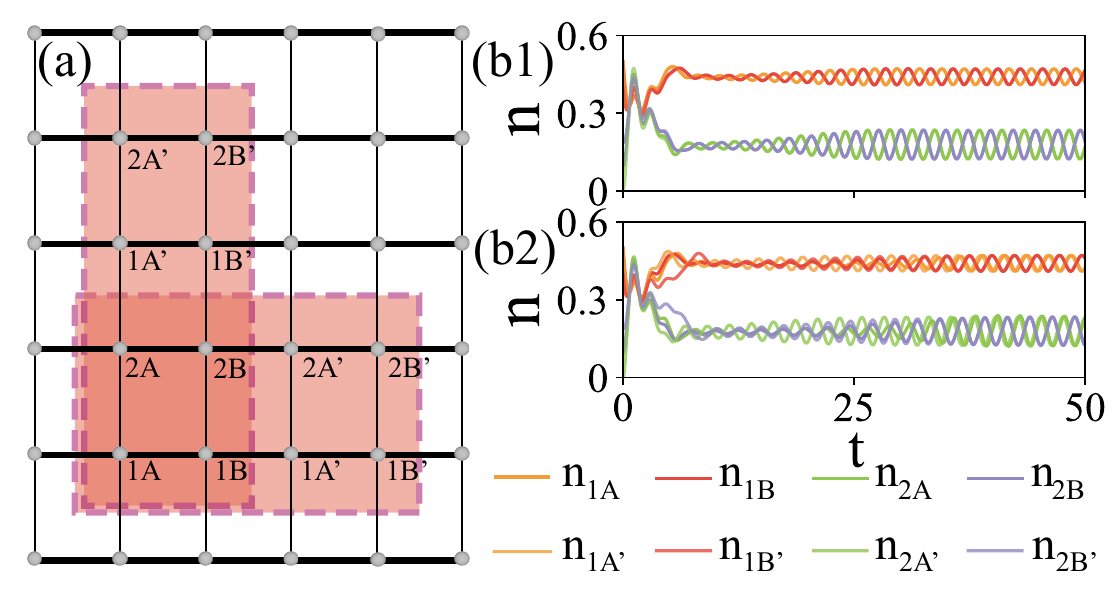}
  \caption{ (a) The sketch map of two-dimensional anisotropy model. The solutions for $4 \times 2$ and $2 \times 4$ rectangle region correspond to figure (b1) and (b2), respectively. (b) Time-evolution for eight sites region by MF approach. The initial Rydberg population for each sites are distinct, and both evolutions finally collapse to $\text{AF}_2$-type oscillation. }
  \label{fig3}
\end{figure}

In this part, we provide the deep understanding to coupled Rydberg chain models and extend it to the more general case. We emphasize that the AF-type perturbation serves as the lowest-order modification of the uniform solutions, and that is the reason why only two sites along the direction of chains, i.e., $A,B$ are included in the MF equation. In other words, for the final solution, the smallest repeating units along both directions are two. Therefore, considering the multiple coupled Rydberg chains with equal spacing [Fig. \ref{fig3}(a)], other non-uniform solutions are unstable, and finally give the solutions with twice lattice spacing period. We verify it in Fig. \ref{fig3}(b) by choosing the larger region for the MF approach. 

In a more general sense, for the two-dimensional and one-sublattice model, only four sites should be included in the MF approximation. For uniform lattice spacing along two directions $V_i = V$, or in other words, uniform two-dimensional distribution of Rydberg density, we could only identify the global AF-type solution \cite{Antiferromagnetic2011Lee}. When considering the anisotropic lattice spacing or interactions, the new synchronization mode will appear on the appropriate parameter scale. Moreover, referring to the stabilization analysis and the numerical results, we infer that except AF-type and $\text{AF}_2$-type oscillations, no other synchronization mode could be found in such systems, and the $\text{AF}_2$-type oscillations could still exist in the presence of long-range interaction. In the language of the time crystal, our findings reveal a phase with coexisting two different continuous time crystals in the two-dimensional Rydberg atomic system.  


\section{Experiment feasibility}

There are two feasible approaches to realize above behavior in the experiment. 

\textbf{Hot Rydberg atomic vapor: }  The time crystal behavior has been observed via EIT in the hot atomic system \cite{Emergence2023Wadenpfuhl,Dissipative2024Wu,Higher2024Liu,Ergodicity2024Ding,Bifurcation2025Liu,Jiao2025Observation}, where the corresponding theoretical description is the limit cycle solution via the MF approximation. Thus in our work, the numerical findings of the time crystal-related behavior should naturally be implemented by two roads of light beam. Adjusting the spacing between two roads should result in the occurrence of the new synchronization mode, which is easily to be distinguished from EIT oscillatory pattern. Nonetheless, there is some uncertainty that the physics of a two-road laser could be well captured by such lattice description, although the single-road case is practicable in both experiment and theory.

\textbf{Rydberg atomic array: }  By comparison, cold atoms have a more precise and controllable parameter setting and lattice system construction, bringing convenience compared to the theoretical phase diagram and experimental setup. The nonlinearity interaction naturally exists between the Rydberg excitations via the blockade effect. The two different oscillatory solutions could be identified by repeatedly taking photos of the Rydberg excitation. However, the conclusion of this paper originates from the MF approach, and we emphasize that there are indeed some differences, compared to the direct solving of the full quantum system.  

\section{Summary and outlook} 

In this work, we discover an emergent synchronization mode in two coupled identical Rydberg atomic chains. By employing the MF approach, we systematically explore the phase properties and the transition points via time-dependent solutions and stability analysis. We find at appropriate choice of the distance between two chains, an emergent oscillatory pattern with respect to the Rydberg density appears, which is distinct from previously reported antiferromagnetic oscillation. We also verify the robustness of such phase against the high-order MF correction, the long-range interaction, demonstrating it as an emergent phase due to the two-dimensional anisotropy. This oscillatory solution is further related to the new pattern of the dissipative time crystal behavior in hot vapor or atomic array systems. In the further work, one could take consideration the feasible experimental condition, such as the higher dimension, external drive, high-order interaction, to construct more complicated and luxuriant time crystal behaviors.

\section{Acknowledgment}

This work is supported by National Natural Science Foundation of China (Project No. 12404275), and the Fundamental Research Program of Shanxi province (Project No. 202403021212015). 

\bibliography{main}

\clearpage

\setcounter{equation}{0}
\setcounter{figure}{0}
\renewcommand{\theequation}{S\arabic{equation}}
\renewcommand{\thefigure}{S\arabic{figure}}
\setcounter{page}{1}
\linespread{1.05}

\begin{widetext}
\begin{center}
\bf \LARGE Supplemental materials of Emergent synchronization mode in coupled Rydberg atomic chains
\end{center}
\end{widetext}

\section*{Mean-field approximation}
\label{sup1}
In this section, we provide the details on the mean-field (MF) approach on the Hamiltonian in Eq. (1) in the main text. As is well-known,  the Master equation in terms of arbitrary observables $O$ writes, 
\begin{equation}
    \begin{aligned}
        \langle \dot O \rangle&= -i \langle \left[ H, O\right] \rangle+ \langle \mathcal{L}(O) \rangle\\
        \mathcal{L}(O) &= \gamma \sum_i \left( \sigma^-_i O \sigma^+_i -\frac{1}{2} \{ n_i, O\} \right).
    \end{aligned}
    \label{eqS1}
\end{equation}
Where $\langle \cdots \rangle$ denotes average value of the physical observables. We consider the diagonal matrix element $n_i$ and the non-diagonal matrix element $\sigma^-_i$ as the key observables. Bringing them into Eq. \eqref{eqS1}, we get, 
\begin{equation}
    \begin{aligned}
        \partial_t \langle n_i \rangle &= \Omega \langle  \text{Im} \sigma_i \rangle - \gamma \langle n_i \rangle \\
        \partial_t \langle \sigma^-_i \rangle  &= -\frac{\gamma}{2} \langle \sigma_i \rangle  - i\Delta \langle \sigma_i \rangle - i\frac{\Omega}{2} (2 \langle n_i \rangle -1) \\
        &+ i \sum_{j} V_{ij} \langle n_j \sigma^{-}_i \rangle 
    \end{aligned}
    \label{eqS2}
\end{equation}
$V_{ij}$ represents the coefficient of the density-density interaction term, hence, $V_{ij} = V, V_i, V_{i2}, \cdots$. The last term in Eq. \eqref{eqS2} comes from the commutator between the interaction term of the Hamiltonian and $\sigma^-_i$. 
\begin{equation}
    \left[ \sum_{jk} V_{jk} n_i n_j, \sigma^-_{i}\right] = i \sum_{j} V_{ij} \langle n_j \sigma^{-}_i \rangle 
\end{equation}
The coupled equation is hard to directly solve due to above nonlinear term, and we could employ the MF approach to transfer it to the linear differential equation. In particular, we adopt the approximation $\langle A B\rangle \approx \langle A \rangle \langle B \rangle$. To give a further simplification, we could consider the translation symmetry of the solution. The simplest case is the uniform solution, with the condition $n = \langle n_i \rangle$ and $\sigma = \langle \sigma^-_i \rangle$. Then the equation of motion (EOM) of the two chains Hamiltonian writes, 
\begin{equation}
    \begin{aligned}
        \partial_t n &= \Omega \text{Im} \sigma - \gamma n \\
        \partial_t \sigma &= -\frac{\gamma}{2} \sigma - i \Delta  n  \sigma - i\frac{\Omega}{2} (2n-1) \\
        &+ i (2V + V_i + 2V_{i2} + \cdots) n \sigma.
    \end{aligned}
    \label{eqS3}
\end{equation}
Note Eq. \eqref{eqS3} returns Eq. 3 in the main text for the single chain. However, Ref. \cite{Antiferromagnetic2011Lee} found that the uniform solution is unstable against antiferromagnetic-type perturbation. Therefore, the proper way is to reduce the symmetry of the solution, in other words, to enlarge the unit cell included in the equation. For single chain case, the stable solution is antiferromagnetic-type with $A,B$ two sublattice contained in the EOM. Generalized to the two coupled chains, we first consider four sublattices, labeled $1A, 1B, 2A, 2B$, as written in Eq. 4 in the main text. 

The solution of Eq. 4 in the main text own the stable mode, so that in coupled chains, no oscillation mode could be found even if we enlarge the unit cell. In the following text, we consider the more general case with multi coupled chains. At this time, we solve the EOM of eight sublattices, which could be enlarged in both directions. Finally, we find no other type stable solutions exists in such system. 

\section*{Other perturbation}

To further explore the robustness of the $\text{AF}_2$-type oscillatory phase, we also provide the numerical calculation on the revised model. By adding the certain perturbation to our original model on Eq. 4 in the main text, we first verify the existence of the Phase 1 mentioned in the main text, where the stable $\text{AF}_2$-type limit cycle appears. Next, we will study its effect on the oscillatory period. Here, we consider two sorts of modifications to the Eq. 4 in the main text: (1) the high-order MF term, (2) the next-nearest neighbor interaction. 

\textbf{The high-order MF term: }
The high-order MF approximation originates from the commutation between the interaction term and the $\sigma^{-}_i$ operator. For the next order, we have following general expression, 
\begin{equation}
    \sum_{j} V_{ij} \langle n_j \sigma^{-}_i \rangle \approx \mathcal{V} \langle n_i \rangle \langle \sigma^{-}_i \rangle + \mathcal{V}_2 \langle n_i \rangle^2 \langle \sigma^{-}_i \rangle + \dots,
\end{equation}
where the coefficient $\mathcal{V}$ contains all the interaction terms related to the site $i$, and $\mathcal{V}_2$ is coefficient of the second-order expansion term. In the above case, $\mathcal{V} = 2V + V_i + 2V_{i2}$. For $\mathcal{V}_2$, a proper assumption is the proportional relation with $\mathcal{V}_2 = r\mathcal{V}_1$, where $r$ is a tunable parameter. 

\begin{figure}[h]
  \centering
  \includegraphics[width=\columnwidth]{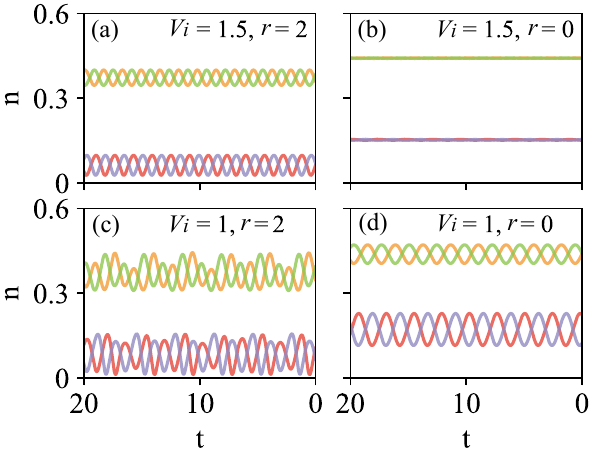}
  \caption{The time-dependent solution of Rydberg population with different $V$ and $r$. All solutions are obtained by the same initial condition.}
  \label{figS1}
\end{figure}

Fig. \ref{figS1} shows the results with $r=2$ as a typical example, compared to the case $r=0$. we observe the complicated oscillatory pattern in (a) that deviates from the sinusoidal form in (b). This is in accordance with the nonlinear effect. Moreover, we  find that at same $V_t=1.5$, the high-order MF term generates the oscillatory phase, which means that nonlinearity renders the larger phase region. 

\begin{figure}[h]
  \centering
  \includegraphics[width=\columnwidth]{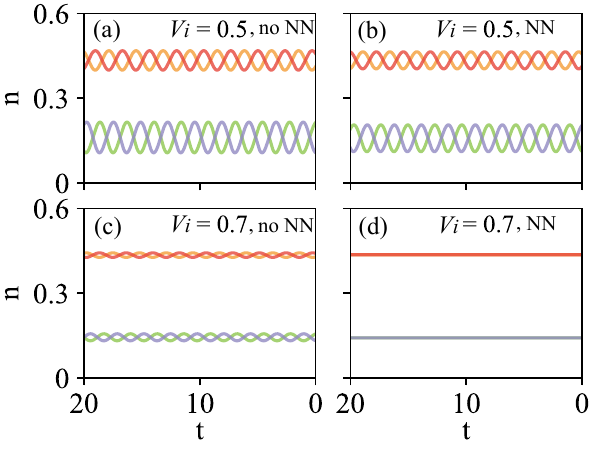}
  \caption{The time-dependent solution of Rydberg population with different $V_t$ and existence of the next-nearest neighbor term for region shape $2 \times 4$. Here, we only display the results of the $1A, 1B, 2A, 2B$ lattice site. In particular for (c,d), we find the Hopf-bifurcation point decreases when we consider the next-nearest neighbor interaction. All solutions are obtained by the same initial condition.}
  \label{figS2}
\end{figure}

\textbf{Next-nearest neighbor interaction: } 
It is known that the long-range interaction term plays the important role in the formation of the emergent phase. Therefore, to study its influence on the $\text{AF}_2$-type oscillation, we include the intra-chain next-nearest neighbor interaction terms $ V_2 n_i n_{i+2}$ in the Hamiltonian. When van der Walls interaction dominates the system, $V_2 = \frac{V}{2^6}$ serves as the small parameter. Since we consider the long-range interaction, we assign the $4 \times 2$ rectangle region to be the unit cell (See Fig. 3 in the main text). The solutions are shown in Fig. \ref{figS2}, which also satisfy $n_{\alpha} = n_{\alpha^\prime}$. We numerically find that the $\text{AF}_2$-type oscillatory range decreases when we consider the next-nearest neighbor interaction. Thus, we conclude that the next-nearest neighbor interaction has a negative effect on the formation of the emergent phase. 

\end{document}